\documentclass[aps,showpacs,twocolumn]{revtex4}
\usepackage[dvips]{graphics,graphicx}
\begin{document}
\title{Semiclassical quantization of the Bogoliubov spectrum}
\author{Andrey R. Kolovsky}
\affiliation{Kirensky Institute of Physics, 660036 Krasnoyarsk,
Russia}
\date{\today }
\begin{abstract}
We analyze the Bogoliubov spectrum of the 3-sites Bose-Hubbard
model with finite number of Bose particles by using a
semiclassical approach. The Bogoliubov spectrum is shown to be
associated with the low-energy regular component of the classical
Hubbard model. We identify the full set of the integrals of
motions of this regular component and, quantizing them, obtain the
energy levels of the quantum system. The critical values of the
energy, above which the regular Bogoliubov spectrum evolves
into a chaotic spectrum, is indicated as well.
\end{abstract}
\maketitle

The Bose-Hubbard model (BH-model),
\begin{equation}
\label{1} 
\widehat{H}= -\frac{T}{2}\sum_{l=1}^{L} \left(
\hat{a}^\dag_{l+1}\hat{a}_l+h.c.\right) +\frac{U}{2}\sum_{l=1}^{L}
\hat{a}^\dag_l\hat{a}^\dag_l\hat{a}_l\hat{a}_l \;, 
\end{equation}
constitutes one of the fundamental Hamiltonians in the condensed
matter theory. The number of phenomena, discussed in the frame of the
BH-model, is so diverse that sometimes it is difficult to see any
link between them. In particular, this concerns the phenomena of
superfluidity and Quantum Chaos. Indeed, the
former phenomenon assumes the regular phonon-like excitation
spectrum, described by the Bogoliubov theory \cite{Land41,Legg01}, while 
the latter phenomenon implies a highly irregular excitation spectrum,
described by the random matrix theory \cite{Cruz90,66,Hill06,70}. This 
seeming contradiction is resolved by noting that these two spectra refer
to different characteristic energies of the system. It is the aim of
the present work to understand of how the regular Bogoliubov spectrum
evolves into an irregular one as the system energy is increased.

To approach the outlined problem we consider the simplest
non-trivial case of the 3-sites BH-model. Indeed, as shown below,
the 3-sites BH-model is well entitled for the Bogoliubov analysis 
and, at the same time, is known to be chaotic. It is worth of noting that
the 3-sites BH-system has been intensively studied during the last decade 
with respect to phenomenon of self-trapping in the system of coupled
nonlinear equations \cite{Eilb85,Henn95}, generalization of the dynamical 
regimes of the celebrated 2-sites BH-system \cite{Nemo00,Fran01}, 
and as a model for multiparticle quantum chaos \cite{Cruz90,Hill06}. 
In the present work we use the 3-sites BH-system as a model for studying 
the Bogoliubov spectrum of the interacting Bose particles in a lattice.

First we discuss the Bogoliubov spectrum of this system, 
following the particularly suited for our aims method of Ref.~\cite{70}. 
The starting point of the analysis is the Hamiltonian of the BH-model 
in the Bloch basis,
\begin{eqnarray}
\label{2} 
\widehat{H}=-T\sum_{k=-1,0,+1} \cos\left(\frac{2\pi
k}{3}\right) \hat{b}_k^\dag\hat{b}_k \\
\nonumber
+\frac{U}{6}\sum_{k_1,k_2,k_3,k_4}
\hat{b}_{k_1}^\dag\hat{b}^\dag_{k_2}\hat{b}_{k_3}\hat{b}_{k_4}
\tilde{\delta}(k_1+k_2-k_3-k_4)  \;,
\end{eqnarray}
which ones obtains from (\ref{1}) by using the canonical
transformation $\hat{b}_k=(1/\sqrt{L})\sum_l \exp(i2\pi
kl/L)\hat{a}_l$. The Hilbert space of (\ref{2}) is spanned by the
quasimomentum Fock states $|n_{-1},n_0,n_{+1}\rangle$, where
$\sum_k n_k=N$ is the total number of particles. To find the
Bogoliubov states of the system (\ref{2}) one uses the anzatz \cite{Legg01}
\begin{equation}
\label{3} 
|\Psi\rangle=\sum_{n=0}^{N/2} c_n |n,N-2n,n\rangle \;.
\end{equation}
Substituting (\ref{3}) in (\ref{2}) and assuming the 
limit $N\rightarrow\infty$, $U\rightarrow0$, $g=UN/L=const$, we
obtain the following equation on the coefficients $c_n$,
\begin{equation}
\label{4} 
2(\delta + g)n c_n + gn c_{n-1}+g(n+1) c_{n+1} = E c_n \;,
\end{equation}
where $g=NU/3$ and $\delta=[1-\cos(2\pi/3)]=1.5$ are the so-called
macroscopic interaction constant and single-particle excitation
energy, respectively. (From now on we set the hopping matrix
element $T$ to unity.) The matrix equation (\ref{4}) can be solved
analytically by mapping it to a differential equation. Namely,
introducing the generating function
$\Phi(\theta)=(1/\sqrt{2\pi})\sum_n c_n e^{i2n\theta}$, we have
\begin{eqnarray}
\label{5a} 
\widehat{H}_{eff}\Phi(\theta)=E\Phi(\theta) \;,\quad
\Phi(\theta+2\pi)=\Phi(\theta) \;,  \\
\label{5b} 
\widehat{H}_{eff}
=(\delta+g)\hat{n}+g\left(e^{-i2\theta}\hat{n}+\hat{n}e^{i2\theta}\right)
\;,\quad \hat{n}=-i\partial/\partial\theta \;.
\end{eqnarray}
It takes only a few lines to prove that the spectrum of the eigenvalue
problem (\ref{5a}) is given by $E_n=-E_{0}+2\Omega n$, where
$\Omega=\sqrt{2g\delta+\delta^2}$ is the Bogoliubov
frequency and $E_{0}=\Omega/2-(\delta+g)$ is the Bogoliubov
correction to the Gross-Pitaevskii ground energy \cite{70}.

It should be mentioned that, instead of (\ref{3}), one can use a
slightly different ansatz, $|\Psi\rangle=\sum_n c_n
|n,N-2n-p,n+p\rangle$. Repeating the above described procedure,
we again obtain a linear spectrum which, however, is now shifted by
$p$ quanta of the Bogoliubov energy with respect to the previous spectrum. 
As a consequence, the excited energy levels of the Bogoliubov spectrum 
are multiple degenerate. To fix the notation we
shall call the number $m$ (which counts the Bogoliubov levels
begging from the ground state $m=0$) by the primary quantum number
and the number $j$, which labels the degenerate sublevels, by the
secondary quantum number. This second quantum number takes 
integer values for even $m$ and half-integer values for odd $m$ in
the interval $|j|\le m/2$. Thus we have
\begin{equation}
\label{6} 
E_{m,j}=\Omega m \;,\quad
\Omega=\sqrt{2g\delta+\delta^2} \;,\quad |j|\le m/2  \;.
\end{equation}
The corresponding to (\ref{6}) wave functions are denoted by
$|\Psi_{m,j}\rangle$ where, by construction,
$|\Psi_{m,j}\rangle=|m/2-j,N-m,m/2+j\rangle$ for $g=0$.

\begin{figure}[b]
\center
\includegraphics[width=8.5cm, clip]{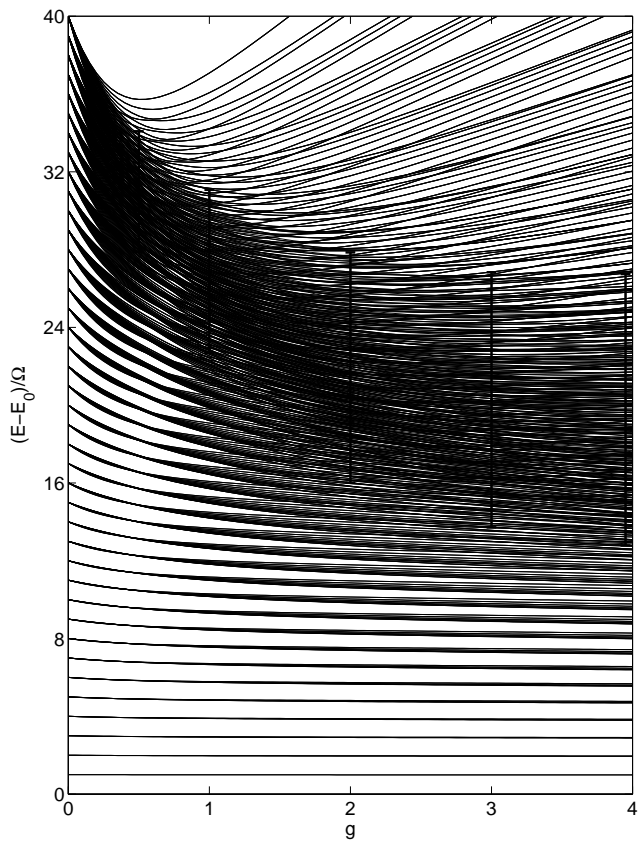}
\caption{Low-energy spectrum of the 3-sites BH-model for $N=40$.
The energy is measured with respect to the ground energy $E_0$ and
scaled with respect to the Bogoliubov frequency $\Omega$. Error
bars indicate the energy intervals, where the classical
counterpart of the system shows chaotic dynamics.} \label{fig1}
\end{figure}

We would like to stress that the above result is valid only in
the limit $N=\infty$ and for any finite $N$ the low-energy
spectrum of the system deviates from (\ref{6}). As an
illustration to this statement, Fig.~\ref{fig1} shows the energy
spectrum of the 3-sites BH-model for $N=40$ and $0\le g\le 4$. It is
seen in the figure that (i) the spectrum is not linear, and (ii)
the Bogoliubov levels are splitted with respect to the secondary
quantum number. In the rest of the paper we shall quantify both of
these effects by using a semiclassical approach. This approach
will also allow us to indicate the critical value of the energy
above which the spectrum of the BH-model is chaotic.

As an intermediate step, let us derive the Bogoliubov spectrum (\ref{6}) by
using semiclassical arguments. The classical counterpart of
the Hamiltonian (\ref{2}) is obtained by scaling it with
respect to the total number of particles, $\widehat{H}/N\rightarrow
H$, and identifying the operators $\hat{b}_k^\dag/\sqrt{N}$ and
$\hat{b}_k/\sqrt{N}$ with pairs of canonically conjugated
variables $(b^*_k,b_k)$. Next we switch to the action-angle
variable, $b_k=\sqrt{I_k}\exp(i\theta_k)$, and explicitly take
into account that $\sum_k I_k=1$ is an integral of 
motion. This reduces our system of 3 degrees of freedom to a
system of two degrees of freedom,
\begin{equation}
\label{7} 
\begin{array}{lll}
H&=&(\delta+g)(I_{-1}+I_{+1}) +2gI_0\sqrt{I_{-1}
I_{+1}}\cos(\theta_{-1}+\theta_{+1})  \\
&-&g(I_{-1}I_{+1}+I_{-1}^2+I_{+1}^2)  \\
&+&2g\sum_{\pm}I_{\mp}\sqrt{I_0
I_{\pm1}}\cos(2\theta_{\mp1}-\theta_{\pm1}) \;,
\end{array}
\end{equation}
where $I_0=1-I_{-1}-I_{+1}$ and the phases $\theta_{\pm1}$ of 
variables $b_{\pm1}(t)$ are measured with respect to the phase of
$b_0(t)$. The low-energy dynamics of the system (\ref{7}), which
is associated with the low-energy spectrum of the system
(\ref{1}), implies $I_{\pm 1}\ll I_0$. Keeping in the Hamiltonian
(\ref{7}) only the terms linear on $I_{\pm 1}$, and using one more
canonical transformation,
\begin{eqnarray}
\label{8}
I=I_{+1}+I_{-1} \;,\quad \theta=(\theta_{+1}+\theta_{-1})/2 \;,\\
\nonumber
J=(I_{+1}-I_{-1})/2 \;,\quad \vartheta=\theta_{+1}-\theta_{-1} \;,
\end{eqnarray}
we obtain
\begin{equation}
\label{9} 
H_{eff}=(\delta+g)I+g\sqrt{I^2-4J^2}\cos(2\theta)  \;.
\end{equation}
[Note that for $J=0$ this Hamiltonian coincides with the classical
counterpart ($\hat{n}/N\rightarrow I$) of the effective
Hamiltonian (\ref{5b}).] Finally, we integrate the system
(\ref{9}) by introducing a new action,
\begin{equation}
\label{10} 
\tilde{I}=\frac{1}{2\pi}\oint I(\theta,\widetilde{E}) d\theta \;,
\end{equation}
and resolving Eq.~(\ref{10}) with respect to the energy. This
gives
\begin{equation}
\label{11} 
\widetilde{E}=\Omega \tilde{I} \;,\quad  
\Omega=\sqrt{2g\delta+\delta^2} \;,\quad \tilde{I}\ll 1 \;.
\end{equation}
Note that the energy is independent of the action $J$, which may be chosen 
arbitrary in the interval $|J|\le \tilde{I}/2$. Referring to the original
quantum problem, this action $J$ obviously labels the degenerate
sublevels of the excited Bogoliubov states \cite{remark0}.

\begin{figure}
\center
\includegraphics[width=8.5cm, clip]{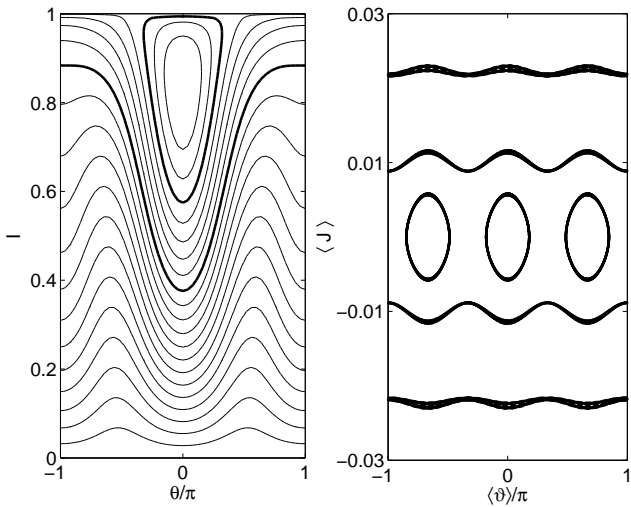}
\caption{Left: phase portrait of the 1D system (\ref{12}) for
$g=1$. The bold lines restrict the chaotic region, where the
trajectories are unstable with respect to variation of
$(J,\vartheta)$. Right: slow dynamics of the variables $\bar{J}=\langle J\rangle$ 
and $\bar{\vartheta}=\langle\vartheta\rangle$ for $g=1$ and $\tilde{I}=0.133$.}
\label{fig2}
\end{figure}

Now we are prepared to discuss the finite-$N$ corrections to the
Bogoliubov spectrum. Let us analyze the dynamics of the classical
system (\ref{7}) in more detail, without assuming $I_{\pm 1}\ll
I_0$. First we shall consider the symmetric solutions, where
$b_{-1}(t)=b_{+1}(t)$. Expressing the Hamiltonian (\ref{7}) in
terms of the canonical variables (\ref{8}) and setting there
$\vartheta=0$ and $J=0$, we have
\begin{eqnarray}
\label{12} 
H_{1D}=(\delta+g)I + g(1-I)I\cos(2\theta) \\
-3gI^2/4 + gI\sqrt{2(1-I)I}\cos(\theta) \;.
\nonumber
\end{eqnarray}
The phase portrait of the 1D system (\ref{12}) is depicted in
Fig.~\ref{fig2}(a). Our particular interest in this phase portrait
are the trajectories near the origin $I=0$, which can be
associated with the Bogoliubov states. It is seen in the figure,
that these trajectories are strongly affected by the elliptic
point in the upper part of the phase space. As a consequence, 
the eigen-frequency of the system depends on the action $\tilde{I}$. 
Namely, $\widetilde{\Omega}=\widetilde{\Omega}(\tilde{I})$ vanishes for 
the separatrix action $\tilde{I}^*$, and for $\tilde{I}\ll\tilde{I}^*$ one has
\begin{equation}
\label{13} 
\widetilde{\Omega}(\tilde{I})=\Omega -\gamma \tilde{I} \;,
\end{equation}
where the nonlinearity $\gamma$ is a unique function of $g$. (For
instance, for $g=0.1,1,4$ we have $\gamma/\Omega=0.1,0.6,1$, respectively.) 
Referring to the original quantum problem the result (\ref{13}) means that 
the energy difference between $(m+1)$-th and $m$-th Bogoliubov levels
decreases as $\gamma m/N$.

\begin{figure}
\center
\includegraphics[width=8.5cm, clip]{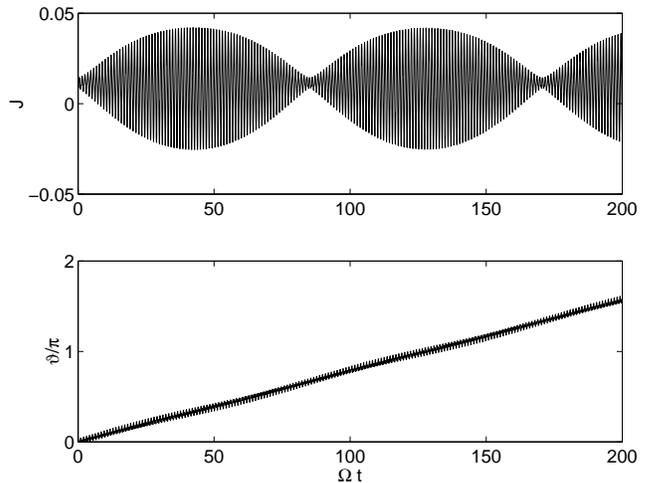}
\caption{An example of time evolution of the conjugated variables
$J(t)$ and $\vartheta(t)$ in the regular regime. Parameters correspond
to the second from top trajectory in Fig.~\ref{fig2}(a).} \label{fig3}
\end{figure}

Next we address the `stability' of the symmetry plane trajectories
depicted in Fig.~\ref{fig2}(a) with respect to variation of $J$.
Within the Bogoliubov approximation the action $J$ is an integral of
motion and may be chosen arbitrary in the interval $|J|\le
\tilde{I}/2$.  It should be understood, however, 
that in reality the action $J$ does depend on time. 
An example of this dependence is given in Fig.~\ref{fig3}.
It is seen that the time evolution of the system is a
superposition of fast dynamics, where $J(t)$ and $\vartheta(t)$
oscillate with the Bogoliubov frequency (more precisely, with the
frequency $\widetilde{\Omega}$), and slow dynamics, with the
characteristic frequency of the orders of magnitude smaller than
the Bogoliubov frequency. Going ahead, we note that this new
frequency defines the splitting of the Bogoliubov levels in
Fig.~\ref{fig1}, and for the moment we only stress that the system
dynamics remains regular. This conclusion holds for any trajectory
of the effective 1D system (\ref{12}), providing that the
trajectory lies well below the separatrix. If we choose a
trajectory closer to the separatrix, we observe a transition from
regular to chaotic dynamics. We identify the exact border of the
transition to chaos by calculating the Poincare cross-section of (\ref{7}) 
for different values of the energy $\widetilde{E}$ and evaluating the volume 
of the chaotic component as the function of energy \cite{remark1}. It is found
that the chaotic region is restricted to a relatively narrow energy interval 
$\widetilde{E}_{min}(g)\le \widetilde{E}\le \widetilde{E}_{max}(g)$.
For $g=1$ the phase trajectories of the system (\ref{12}), corresponding 
to $\widetilde{E}_{min}$ and $\widetilde{E}_{max}$, are marked
by the bold lines in Fig.~\ref{fig2}(a).
Additionally, the error bars in Fig.~\ref{fig1} indicate the
chaotic energy intervals for different values of $g$. The depicted
borders are consistent with the visual analysis of the spectrum
and suggest the following simple criteria of the transition to
chaos: it takes place when the total splitting of the Bogoliubov levels
with respect to the second quantum number exceeds the mean
distance $\widetilde{\Omega}$ between the levels.

The question on the sublevels splitting is in turn. As it was
already mentioned, this splitting is defined by the slow dynamics
of the system. To address this slow dynamics we introduce the new
variables $\bar{J}=(1/2\pi)\oint J d\tilde{\theta}$ and
$\bar{\vartheta}=(1/2\pi)\oint \vartheta d\tilde{\theta}$, where
$\tilde{\theta}=\widetilde{\Omega} t$ is the phase
conjugated to the action $\tilde{I}$. (Note that the action
$\tilde{I}$ is an adiabatic integral of motion and, hence,
does not depend on time.) The Hamiltonian
equations of the motion for the variables $\bar{J}$ and
$\bar{\vartheta}$ read,
\begin{equation}
\label{14}
\begin{array}{lllll}
\dot{\bar{\vartheta}}& =&\langle \partial H/\partial\vartheta\rangle
&\approx&-2g\bar{J} \;, \\
\dot{\bar{J}}&=&-\langle \partial H/\partial J\rangle 
&\approx& 6g V(g,\tilde{I})\sin(3\bar{\vartheta}/2) \;,
\end{array}
\end{equation}
where $V(g,\tilde{I}) \approx \langle(I/2)^{3/2}\cos{\theta} \rangle$ and 
$\langle\ldots\rangle$  means time average over one period of the fast dynamics. 
Thus the slow dynamics is defined by the pendulum-like Hamiltonian,
\begin{equation}
\label{15} H_{slow}=-g\bar{J}^2 +
4gV(g,\tilde{I})\cos(3\bar{\vartheta}/2) \;.
\end{equation}
It is worth of noting that, to obtain (\ref{15}), we have assumed
the quantity  $V(g,\tilde{I})$ to be independent of $\bar{J}$,
which can be justified only if $\bar{J}\ll\tilde{I}/2$. Nevertheless, 
the Hamiltonian (\ref{15}) is found to well capture the main features of the
low-energy regular dynamics for arbitrary $\bar{J}$. In
particular, it correctly predicts the existence of stable points
at $\vartheta=0,\pm 4\pi/3$, where the phases of  $b_{\pm 1}(t)$
are locked to 0 and 120 degrees with respect to each other [see
Fig.~\ref{fig2}(b)]. The size of the stability islands around
these fixed points is obviously given by the separatrix trajectory
of the pendulum, i.e., is proportional to $|V(g,\tilde{I})|^{1/2}$.

It is instructive to consider the limiting case $g\rightarrow0$.
As easy to show, in this limit $V(g,\tilde{I})\rightarrow0$ and,
hence, $H_{slow}=-g\bar{J}^2$. Let us prove that this results
corresponds to the first order quantum perturbation theory on $U$.
Indeed, calculating the first order corrections to the energies of
the quasimomentum Fork states
$|\Psi_{m,j}\rangle=|m/2-j,N-m,m/2+j\rangle$, we have $\Delta
E=(U/2L) \langle\Psi_{m,j}| \sum
 \hat{b}_{k_1}^\dag\hat{b}^\dag_{k_2}\hat{b}_{k_3}\hat{b}_{k_4}
\tilde{\delta}(k_1+k_2-k_3-k_4) |\Psi_{m,j}\rangle \sim -(U/L)
j^2$, or $\Delta\widetilde{E}=-g\bar{J}^2$, where $\bar{J}=j/N$
and $\widetilde{E}=E/N$. Thus for small $g$ the splitting between 
sublevels grows linearly with $g$. This linear regime changes to 
a nonlinear one as soon as the second term in the Hamiltonian (\ref{15}) 
takes a non-negligible value. This second term also causes the rearrangement 
of the sublevels, clearly seen in Fig.~\ref{fig1}. Needless to say that 
in this case the second quantum number is defined by the action 
$\tilde{J}=(1/2\pi)\oint\bar{J}d\bar{\vartheta}$, which amounts to 
the phase volume encircled by the trajectories in Fig.~\ref{fig2}(b).

We conclude the paper by formulating quantitative criteria for
the onset of Quantum Chaos. As mentioned earlier, the transition
to irregular spectrum occurs when the total splitting of the $m$-th Bogoliubov 
level compares with the Bogoliubov frequency. Ignoring the nonlinear corrections, 
one has $g(m+1)^2/4N\sim \Omega$, or
\begin{equation}
\label{16} 
m_{cr}\approx\left(\frac{4N\sqrt{\delta^2+2\delta g}}{g}\right)^{1/2} \;.
\end{equation}
Through the relation $E_{cr}\approx \Omega m_{cr}$ this estimate defines 
the critical value of energy above which the regular spectrum transforms 
into a chaotic one.

In conclusion, we have analyzed the Bogoliubov spectrum of the BH-model. 
This spectrum corresponds to low-energy excitations of the system and is 
usually introduced by using  the Bogoliubov-de Gennes transformation. 
This standard method, however, is rather formal and hides the underlying 
classical dynamics of the BH-model. In this work we use a semiclassical 
method which, by definition, explicitly refers to the classical dynamics 
and provides in this way a deeper insight in the structure of the low-energy 
spectrum of the system. In the present work we restricted ourselves by 
considering the 3-sites BH-model, although many of the reported results 
hold for $L>3$ as well. An advantage of the 3-sites model is that, thanks to 
a relative low dimensionality of the system, its classical dynamics can be 
understood in every detail. In particular, the phase space of the system 
essentially consists of two regular and one chaotic component in between, 
where the low-energy regular component is shown to be associated with 
the Bogoliubov spectrum. We identify the full set of the integrals of motion 
for this low-energy regular component and, quantizing them, obtain the 
low-energy levels of the quantum BH-model. These levels are labelled by 
two quantum numbers, $m$ and $j$. The first quantum number $m$ corresponds 
to usual Bogoliubov ladder, where the distance between neighboring levels 
is approximately given by the Bogoliubov frequency $\Omega$ 
(i.e., $E_{m+1,j}-E_{m,j}\sim\Omega$). The second quantum number $j$ labels 
$(m+1)$ sublevels of the $m$-th Bogoliubov level, where the splitting between 
the sublevels is proportional to the interaction constant $g$ and inverse 
proportional to the system size $N$ (i.e., $E_{m,j+1}-E_{m,j}\sim g/N$). 
If we go up the energy axis, the total splitting of the Bogoliubov levels 
compares the distances between the levels and the energy spectrum shows 
a transition from a regular to irregular (chaotic) one.

The described scenario of evolution of the Bogoliubov spectrum into a regular, 
Bogoliubov-like spectrum and further into a chaotic spectrum also holds for 
the BH-model with $L>3$ sites. However, to indicate the critical energies for 
these transitions remains  an open problem. The qualitative difference between 
the 3-sites and, for example, 5-sites BH-models is that the latter system has 
two different Bogoliubov frequencies, associated with two different single-particle 
excitation energies. For some values of the macroscopic interaction constant $g$, 
these frequencies become commensurable, which strongly affects the onset of chaos. 
We reserve this problem of interacting Bogoliubov spectra for future studies.

Fruitful discussions with S.~Tomsovic and partial support by Deutsche 
Forshungsgemeinschaft (within  the SPP1116 program) is gratefully acknowledged.



\begin{thebibliography}{10}
\bibitem{remark0}
Since $1/N$ plays the role of Planck's constant in the considered problem,
the semiclassical quanitization of (\ref{12}) corresponds to
$\tilde{I}=m/N$, $m=0,1,\ldots$, and $J=j/N$, $j=-m/2,-m/2+1,\ldots,m/2$.
\bibitem{remark1}
There is some freedom in defining the border of the transition to chaos.
In what follows we define it as the energy below which the chaotic component
occupies less than one percent of the energy shell.
\end{thebibliography}
\end{document}